\documentclass[pra,twocolumn,showpacs,superscriptaddress,floatfix]{revtex4}

\usepackage{graphicx}
\usepackage{dcolumn}
\usepackage{amsmath,amsfonts,bm}

\begin{document}
\title{Raman Coupling of Zeeman Sublevels in an Alkali Bose Condensate}

\author{K. C. Wright}
\affiliation{Department of Physics and Astronomy, University of Rochester, Rochester, NY 14627}
\author{L. S. Leslie}
\affiliation{Institute of Optics and Laboratory for Laser Energetics, University of Rochester, Rochester, NY 14627} 
\author{N. P. Bigelow}
\affiliation{Department of Physics and Astronomy, University of Rochester, Rochester, NY 14627}
\affiliation{Institute of Optics and Laboratory for Laser Energetics, University of Rochester, Rochester, NY 14627}         

\date{\today}  

\begin{abstract}
We investigate amplitude and phase control of the components of the spinor order parameter of a $^{87}$Rb Bose-Einstein condensate.  By modeling the interaction of the multilevel atomic system with a pair of Raman-detuned laser pulses, we show that it is possible to construct a pulse-sequence protocol for producing a desired state change within a single Zeeman manifold.  We present several successful elementary tests of this protocol in both the $F=1$ and $F=2$ Zeeman manifolds of $^{87}$Rb using the D$_1$ transitions.   We describe specific features of the interaction which are important for multimode, spatially-varying field configurations, including the role of state-dependent, light-induced potentials.  
\end{abstract}

\pacs{ 32.60.+i, 03.75.Mn, 42.50.Ex}

\maketitle

\section{Introduction}
Dilute gases of alkali atoms provide a unique setting in which to study the properties of quantum fluids. The tremendous precision with which they can be controlled and probed makes them an ideal model system for understanding physical phenomena important to other less easily studied systems with similar physical properties.  Because alkali atoms possess internal spin degrees of freedom, alkali BECs are described by a multi-component order parameter.  This vectorial nature of the order parameter allows for much more complex structure and dynamics than occur in a single component BEC, with important ramifications for the system's physical properties.  The work contained in this Article was motivated by an interest in creating and studying topological excitations of the spinor BEC such as coreless vortices \cite{OhmiJPSJ98,HoSpinorPRL98}.  Spin textures such as coreless vortices \cite{AndersonPhasePRL77, MerminCirculationPRL76,IsoshimaQuantumJPSJ01,MizushimaMermin-HoPRL02,IsoshimaAxisymmetricPRA02} and skyrmions \cite{AlKhawajaSkyrmionsN01,BattyeStablePRL02} are important features of quantum many-body systems such as liquid helium \cite{FeynmanApplicationPiLTP27, HallRotationAIP60}, superconductors \cite{BlatterVorticesRoMP94}, and neutron stars \cite{BaymSuperfluidityN69}, and they also possess important symmetry properties deeply connected to fundamental principles of physics \cite{HoLocalPRL96}. 

There have been several previous approaches to creating nontrivial spin textures in alkali BECs.  Using a method referred to as `phase imprinting', the first vortex observed in an alkali BEC was created by coupling hyperfine ground states using a combination of laser, radio-frequency, and microwave fields \cite{WilliamsPreparingN99,MatthewsVorticesPRL99,AndersonWatchingPRL01}.  While this was quite effective, the interaction time required to generate the desired state was on the order of tens of milliseconds.  In another approach,  Raman transitions using EIT-based stopped light techniques \cite{DuttonObservationS01} were used to generate solitons and vortices.  Both this approach and the phase imprinting  experiments involve population transfer between different hyperfine (ground) spin states, which are typically separated by several GHz.  This large energy splitting essentially freezes out spin dynamics.  In a elegant and very different series of experiments,  adiabatic ramping of an inhomogeneous magnetic field was shown to create spin textures within a single Zeeman manifold \cite{StengerSpinN98,LeanhardtImprintingPRL02,LeanhardtCorelessPRL03,SadlerSpontaneousN06}. However, the spatial complexity of the spin textures that can be created using purely magnetic techniques is restricted by practical constraints on creating complex field geometries and by conservation of angular momentum within the essentially closed system.

In this work we explore the interaction of Raman-detuned laser fields with an alkali BEC, with the purpose of developing general all-optical techniques for controlling the multi-component order parameter \cite{WrightOpticalPRA08}.  We choose an all optical approach because state manipulation can be readily realized on a microsecond time scale, and we focus on spin states within a single hyperfine manifold because this provides full access to spin mixing dynamics \cite{LawQuantumPRL98}. Because of the true multilevel nature of the alkali atoms, the atom-field interaction exhibits a remarkable degree of complexity which, as we will show, must be taken into account. An interest in multilevel Raman transitions of this type is not new.  In earlier work, for example, the coupling between three separate electronic states with Zeeman substructure has been studied  \cite{MartinCoherentPRA96,MartinCoherentPRA95,ShoreCoherentPRA95}.  Further, a number of elegant experiments showing coherent control and creation of superpositions of Zeeman sublevels has been conducted with atomic beams of metastable neon \cite{HeinzPhaseOC06,VewingerAmplitudeIPRA07,VewingerAmplitudeIIPRA07}.   We also note that the idea of using a stimulated Raman process to couple multiple ground-state Zeeman sublevels was previously explored theoretically \cite{LawSynthesisOE98} and that adiabatic passage between ground state sublevels of a cesium atomic beam has been demonstrated experimentally \cite{PilletAdiabaticPRA93}.  However, to our knowledge, our results here represent the first detailed treatment of Raman coupling between sublevels of the same ground-state Zeeman manifold in a way that takes into account both the excited state and the ground state level structure.

One of the more interesting and useful features of the Raman coupling interaction investigated here is that the applied laser fields induce state-dependent light shifts which depend in a non-trivial way on the laser beam parameters and the atomic level structure \cite{MathurLightPR63,TannoudjiExperimentalPRA72,ChangMeasurementPRA01}.  This inherent richness in the interaction provides a convenient degree of flexibility in controlling the condensate, allowing the creation of a variety of states not accessible in previous experiments.  In Section \ref{sec:modeling} we outline the construction of a model of the system, subject to certain constraints which are of particular importance in modeling spatially inhomogeneous interactions \cite{WrightOpticalPRA08}.  Although the results presented here can be easily extended to other alkali atoms, we will focus on implementation in the $F=1$ and $F=2$ ground state sublevels of $^{87}$Rb, which has a number of features that recommend it as a good choice for investigating BEC spinor physics. 

In Section \ref{sec:results} we show the experimental results of several tests of the validity of the model in $^{87}$Rb as put forward in Section \ref{sec:modeling}.  These tests include demonstrations of amplitude and phase control of the components of the order parameter via pair-wise coupling, as well as a demonstration of the possibilities for simultaneous three-state coupling in the $F=2$ manifold.  The application of these results to the broader problem of preparing specific spin textures with spatially-varying amplitudes and phases will be briefly discussed in the conclusion.

\section{Modeling the Raman-coupled Spinor BEC}\label{sec:modeling}
The $^2S_{1/2}$ ground electronic state in alkali atoms is composed of two hyperfine manifolds of total spin $F=I\pm 1/2$, where $I$ is the nuclear spin.  These manifolds consist of $2F+1$ sublevels, which in the absence of a magnetic field are degenerate.  These states are connected to the $^2P_{1/2}$ electronic states by the D$_1$ transitions, which in the case of Rb involves photons of $\lambda=795$ nm. The Land\'{e} factors of the ground and excited electronic states differ; in the presence of a small magnetic field the transitions between individual sublevels are no longer degenerate.  In constructing an appropriate model of our experimental configuration, we consider the presence of a small applied magnetic bias field of up to a few tens of Gauss.  Because the two-photon linewidth of the Raman coupling can easily be smaller than the nonlinear part of the Zeeman shift (see Section \ref{ssec:2pdet}), it is necessary to use the Breit-Rabi formula to calculate the magnitude of the energy shifts due to the applied magnetic fields.  This has especially important consequences when linking three or more states in the same Zeeman manifold. 

In both our modeling and our experiments we work in the low-density limit, which has two important consequences.  First, the nonlinear interaction terms due to the mean-field energy are small enough to be neglected in comparison to the Raman coupling terms.  Second, in a sufficiently low-density atomic cloud, beam propagation effects can be safely neglected.  We also adopt a pure state description of the system, which is valid as long as the spontaneous scattering rate is negligible.  This can be accomplished either by detuning the optical fields sufficiently far from resonance, or by using laser pulse configurations which maintain the system in an adiabatically evolving dark state (i.e. STIRAP) \cite{MartinCoherentPRA96, MartinCoherentPRA95, ShoreCoherentPRA95}.  Spontaneous heating of the condensate is undesirable in any case, therefore we are required to work in a regime where the photon scattering rate is small regardless of the state description. 

\begin{figure}[ht]
	\centering
		\includegraphics[scale=1]{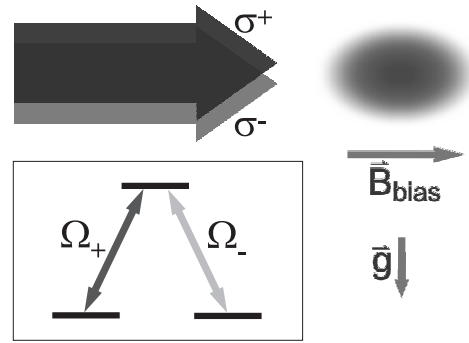}
	\caption{Experimental geometry indicating the relative orientation of the Raman beams, the magnetic quantization axis, and gravity.  The inset is a schematic indicating the linkage pattern characteristic of the $\sigma^+$ - $\sigma^-$ coupling configuration.}
	\label{fig:geom}
\end{figure}

The beam interaction geometry we have chosen is shown in Fig \ref{fig:geom}.  The two Raman beams are copropagating, parallel to the quantization axis defined by the magnetic bias field.  They are $\sigma^+$ and $\sigma^-$ polarized so that coupling occurs between states with $\Delta m_F=2$.   Because the beams are collinear, the atoms' change in kinetic energy due to the Raman transition is many orders of magnitude smaller than the condensate momentum distribution.  This makes it possible to drop the kinetic energy terms from the Hamiltonian. It is also important to note that after the optical interaction the populations in different internal states have no significant relative momentum, and remain physically overlapping after the interaction has taken place. 

Although we have limited the model and results presented here to the $\sigma^+$ and $\sigma^-$ coupling configuration, it should be possible to extend these results to include $\pi$ transitions.  This would allow all of the Zeeman sublevels in a manifold to be accessed, instead of the subset corresponding to $\Delta m_F=2$. By using beam modes with polarization components in the direction of propagation, this can be accomplished while still using a collinear beam geometry, i.e. by driving anomalous Raman transitions \cite{MarzlinCreationPRA00}. Experimental work exploring this intriguing possibility is in progress.

One important feature we wish to highlight in the model presented here is the consideration of complex-valued Rabi frequencies.  For interactions involving plane waves it is customary to constrain the Rabi frequencies to be real valued by factoring out an overall phase.  For multimode configurations where the relative phase of the Raman beams is allowed to vary spatially, the Rabi frequencies cannot be treated as real valued over the entire interaction region.  The experimental results presented in Section \ref{sec:results} involve spatially uniform interactions, and so do not address this additional degree of freedom in the system, however the implications for multimode interactions will be discussed in the conclusions.

In the theory and experiments described in this Article, the detunings of the laser fields have been considered to lie within a range spanning about twice the excited state hyperfine splitting of 816 MHz.  We will show that for the $^{87}$Rb D$_1$ transitions there is a wide range of of intensities and detunings for which the coherent two-photon effects dominate and the photon scattering rate is negligible. In addition, we limit our consideration to Rabi frequencies much smaller than the excited state hyperfine splitting, also noting that counter-rotating terms in the Hamiltonian are typically far enough from resonance with any real transitions that they can be neglected.  For alkali atoms with smaller hyperfine splittings, such as sodium, resonant excitation places tighter limits on the range of detunings that are experimentally useful.

Bearing in mind all of these considerations, one can write a Hamiltonian for the system in the interaction picture \cite{ShoreTheory90}, apply the rotating wave approximation, and adiabatically eliminate the excited states \cite{FewellAdiabaticOC05}.  This procedure gives a greatly simplified effective Hamiltonian for the system which is valid and useful for most interaction configurations of experimental interest.  We will investigate the application of this approach to the $F=1$ and $F=2$ ground state hyperfine manifolds of $^{87}$Rb in the sections that follow.

\subsection{ Application to the $^{87}$Rb $F=1$ Manifold}\label{ssec:F=1}

\begin{figure}[ht]
	\centering
		\includegraphics[scale=1.1]{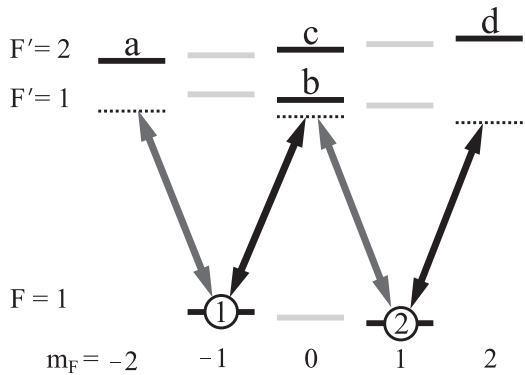}
	\caption{State linkage diagram for $(\sigma^+,\sigma^-)$ Raman coupling of the $F=1$ ground state manifold of $^{87}$Rb via the D$_1$ transitions. The magnitude of the Zeeman shift of the states by the small magnetic bias field is exaggerated for clarity.}
	\label{fig:F1levels}
\end{figure}

The state linkages for the system composed of the $F=1$ Zeeman manifold coupled to the D$_1$ excited states are shown in Fig.~\ref{fig:F1levels}  Applying the procedure outlined above results in an effective two-level system, with coupling between the two indicated Zeeman sublevels described by a Hamiltonian with the following form.

\begin{equation}\label{eq:F=1Ham}
\hbar
\left[ \begin{array}{ccc} 
\frac{\delta}{2}+\chi_1^a I_{-} + \chi_1^{bc} I_{+}
&  
\eta_{12}^{bc}\sqrt{I_-I_+}e^{-i\xi_{\pm}}                         
\\ &
\\
\eta_{12}^{bc}\sqrt{I_-I_+}e^{i\xi_{\pm}}
& 
-\frac{\delta}{2} + \chi_2^d I_{+} + \chi_2^{bc} I_{-}
\end{array}\right]
\end{equation}

\noindent $I_+$ and $I_-$ are the intensities of the $\sigma^+$ and $\sigma^-$ polarized beams, and $\delta$ is the two-photon detuning defined below.  The sub(super)scripts refer to ground (excited) states indicated by the labels in the state linkage diagram in Fig.~\ref{fig:F1levels}.  The parameters $\chi$ are coupling coefficients for the state-dependent light shifts.  Their definitions are as follows.

\begin{equation}\label{eq:F1chis}
\begin{split}
	\chi_1^a   &=-\frac{d_{D_1}^2 }{2c\epsilon_0} \left(\frac{(C_1^a)^2}{\Delta^{(a)}} \right)  \\
  \chi_1^{bc}&=-\frac{d_{D_1}^2 }{2c\epsilon_0} \left(\frac{(C_1^b)^2}{\Delta^{(b)}} + \frac{(C_1^c)^2}{\Delta^{(c)}}\right)  \\
  \chi_2^{bc}&=-\frac{d_{D_1}^2 }{2c\epsilon_0} \left(\frac{(C_2^b)^2}{\Delta^{(b)}} + \frac{(C_2^c)^2}{\Delta^{(c)}}\right)  \\
  \chi_2^d   &=-\frac{d_{D_1}^2 }{2c\epsilon_0} \left(\frac{(C_2^d)^2}{\Delta^{(d)}} \right)   
\end{split}
\end{equation}

\noindent The quantity $d_{D_1}$ is the reduced dipole matrix element for the $^{87}$Rb D$_1$ transitions, the $C_i^\alpha$ are the Clebsch-Gordan coefficients for the specific transition between ground state $i$ $(=1,2)$ and excited state $\alpha$ $(=a,b,c,d)$.  The $\Delta$ are detunings defined as follows.

\begin{align*}
2\Delta^{(a)} &=(2E_a-E_1-E_2)/\hbar-( 3\omega_- -\omega_+ )     \\
2\Delta^{(b)} &=(2E_b-E_1-E_2)/\hbar-(\omega_+ + \omega_-)     \\
2\Delta^{(c)} &=(2E_c-E_1-E_2)/\hbar-(\omega_- + \omega_+)     \\
2\Delta^{(d)} &=(2E_d-E_1-E_2)/\hbar-( 3\omega_+ -\omega_-)
\end{align*}

\noindent The $E_n$ represent the atomic bare state energies, which depend on the magnitude of the magnetic bias field.  We have also defined a two-photon detuning

\begin{equation*}
\delta=(E_1 - E_2)/\hbar + (\omega_+ -\omega_-)
\end{equation*}

\noindent The parameter $\eta$ in the Hamiltonian is a coupling coefficient for the effective two-photon Rabi frequency. 

\begin{equation}\label{eq:F1eta}
\eta_{12}^{bc}=-\frac{d_{D_1}^2 }{2c\epsilon_0} \left(\frac{C_1^b C_2^b}{\Delta^{(b)}} + \frac{C_1^c C_2^c}{\Delta^{(c)}}\right)e^{-i\xi_{\pm}}
\end{equation} 

For completeness we have explicitly included a complex exponential term containing  the difference in phase between the $\sigma^+$ and $\sigma^-$ beams, represented as the value $\xi$.  As mentioned in the previous section, allowing the Rabi frequencies to take on complex values is necessary for modeling interactions where the relative phase may vary in space or change as a function of time.

\subsubsection{Pseudospin Representation and Fictitious Fields}

Having determined the form of the system Hamiltonian, it is instructive to rewrite it in a basis of spin-1/2 operators.

\begin{equation}\label{eq:F1spinham}
\hat{H}=\frac{\tilde\alpha}{2}\:\mathbb{I} + \frac{\tilde\delta+\delta}{2}\:\bm{\sigma}_{z} + \cos(\xi_{\pm})\frac{\tilde\Omega}{2}\:\bm{\sigma}_{x} + \sin(\xi_{\pm})\frac{\tilde\Omega}{2} \:\bm{\sigma}_{y}
\end{equation}

\noindent Here we have defined a scalar light shift $\tilde\alpha$, Zeeman light shift $\tilde\delta$, and two-photon Rabi freqency $\tilde\Omega$.

\begin{align}
\tilde\alpha&=I_-(\chi_1^a+\chi_2^{bc})+I_+(\chi_1^{bc}+\chi_2^d) \label{eq:alpha}\\ 
\tilde\delta&=I_-(\chi_1^a-\chi_2^{bc})+I_+(\chi_1^{bc}-\chi_2^d)\label{eq:tdelta}\\
\tilde\Omega&=2\:\eta_{12}^{bc}\sqrt{I_+I_-} \label{eq:tOmega}
\end{align}

Reconstructing the Hamiltonian in the form of Eq. \ref{eq:F1spinham} gives additional insight into the expected behavior of the system, allowing us to visualize it as a spin-1/2 system subject to a fictitious electric and magnetic field \cite{MathurLightPR63, TannoudjiExperimentalPRA72}. The scalar term involving $\tilde\alpha$ results in a global phase shift that is trivial for uniform laser fields, and has no effect on the internal state evolution.  It should be noted, however that for laser fields with spatially varying intensities, the energy shift it represents can be physically important, for example it contributes to the dipole force acting on the atoms.  The remaining terms involving spin operators and the parameters $\tilde\delta$, $\tilde\Omega$, and $\xi$ can be conveniently understood as an interaction with a fictitious magnetic field.

\begin{equation}
\bm{B}_f = \frac{\tilde\Omega \cos(\xi_{\pm}) \bm{\hat{\imath}}+\tilde\Omega\sin(\xi_{\pm})\bm{\hat{\jmath}} +(\tilde\delta+\delta)\bm{\hat{k}}}{\mu_B/ \hbar}
\end{equation}

 If only one or the other of the laser fields is applied, $\tilde\Omega=0$, and the fictitious magnetic field is oriented parallel or antiparallel to the polar axis. Assuming the intensity-independent, two-photon detuning $\delta$ is zero, the magnitude and sign are determined by that of the Zeeman light shift parameter $\tilde\delta$.  
 
\begin{figure}[ht]
	\centering
		\includegraphics[scale=1.0]{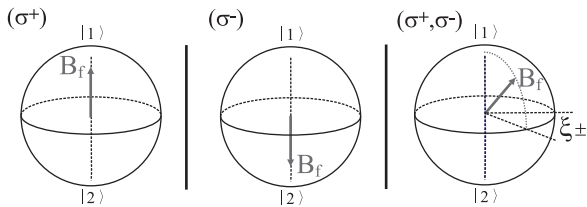}
	\caption{Representation of the fictitious magnetic field due to the application of the $\sigma^+$ laser field (left), the $\sigma^-$ field (center), and both $\sigma^+$ and $\sigma^-$ fields together (right). For these plots, the detuning is assumed to be midway between the $F'=1$ and $F'=2$ excited state manifolds.}
	\label{fig:fictBsphere}
\end{figure} 
 
When both laser fields are turned on, $\tilde\Omega$ is nonzero, and the orientation of the fictitious magnetic field is no longer solely along the polar axis.  The polar angle is determined by the relative magnitudes of the laser field intensities $I_+$ and $I_-$.  The azimuthal angle of the fictitious magnetic field depends on the relative phase of the fields $\xi_\pm$. The relationship between the laser fields and the orientation of the fictitious magnetic field in the pseudospin space is shown in Fig.~\ref{fig:fictBsphere} for the specific case where $\Delta_{D_1}=0$.

\begin{figure}[ht]
	\centering
		\includegraphics[scale=1.0]{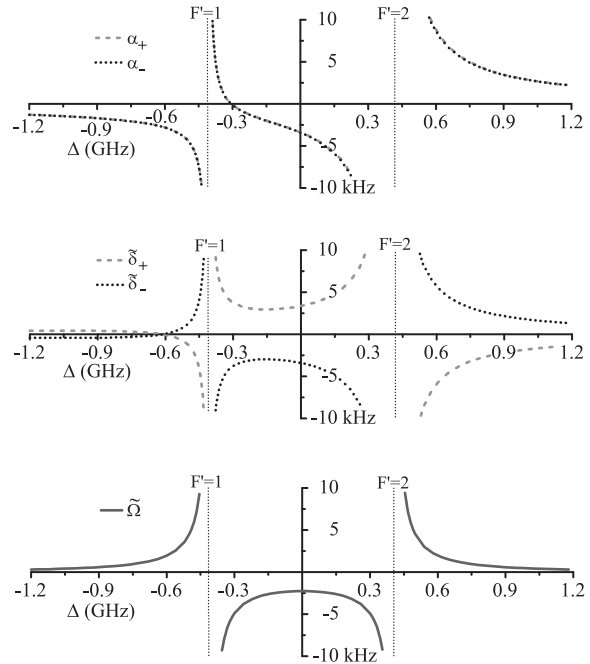}
	\caption{Detuning dependence of the scalar light shift $\tilde\alpha$, Zeeman light shift $\tilde\delta$, and two-photon Rabi frequency $\tilde\Omega$.  The scalar and Zeeman light shifts are expressed in terms of their components which depend on intensities of the $\sigma^+$ and $\sigma^-$ polarized laser fields. ($I_+,I_-=1$ mW/cm$^2$, B = 1.33 Gauss) }
	\label{fig:alphadeltaOmegaF1}
\end{figure}

As can been seen from Fig.~\ref{fig:alphadeltaOmegaF1}, the dependence of the parameters $\tilde\alpha$, $\tilde\delta$, and $\tilde\Omega$ on detuning from the excited states is nontrivial.  In the figure the parameters $\tilde\alpha$ and $\tilde\delta$ are shown broken down into the parts which depend on the $\sigma^-$ and $\sigma^+$ polarized fields. It should be noted that these plots are valid as long as the detuning from resonance is sufficiently large compared to the Rabi frequencies for the involved transitions.  For intensities typical of our experiments ($<$70 mW/cm$^2$), the plots are valid except in a region less than 100 MHz wide around the resonances shown.  The complexity in these curves clearly indicates that the detuning from the excited state must be taken into consideration when choosing how to act on the system with the laser pulses.  In actual practice, we have found that this feature affords a degree of control over the interaction Hamiltonian that can be highly useful.

\subsubsection{Determining a Pulse Protocol}

The motivation for the analysis of the preceeding section is to determine how to control the state of the system, and cause it to evolve in a desired manner.   Because the general picture of how a spin evolves in a specified external field is well understood, this representation is of great utility in understanding how to approach the problem.  A number of advanced spin manipulation techniques have been developed for NMR which could potentially be implemented in this system \cite{WimperisBroadbandJMR94}, however we focus here on using simple pulse configurations. 

Adiabatic passage techniques (i.e. STIRAP) can be used to perform coherent transfer \cite{WrightOpticalPRA08}, but are not well suited for certain applications. For example, the large pulse areas required for efficient transfer make it comparatively difficult to compensate for the inhomogeneous light shifts which arise when using beam modes with spatially varying relative intensities.  We will not address the use of adiabatic techniqes here, focusing instead on coherent population oscillation driven with square pulses. Specifically, we assume the use of sequences of laser pulses of constant intensity $I_-$, $I_+$, and duration $\tau$.  We then determine what laser fields to apply in order to change the polar or azimuthal orientation of the pseudospin vector representing the state of the system. 

From \eqref{eq:tdelta} and \eqref{eq:tOmega} we see that in order to cause complete Rabi oscillations between the two states at a frequency $\tilde\Omega$, we must turn on both fields using a fixed ratio of intensities that makes $\tilde\delta+\delta=0$.  In the pseudospin space, this corresponds to generating a fictitious magnetic field which lies in the equatorial (x-y) plane, such that the pseudospin vector will precess in a circular path between $\left|1\right\rangle$ and $\left|2\right\rangle$ at the antipodes.  The change in the polar angle $\theta$ of the pseudospin is in that case simply

\begin{equation}\label{eq:thetaconst1}
\theta =\tilde\Omega \: \tau=2 \eta_{12}^{bc}\sqrt{I_-I_+}\tau\\
\end{equation}

\noindent where $\tau$ is the pulse duration.   It is important to note, however, that because the sign and magnitude of the light shifts depend in a complicated way on the atomic level structure and the detuning, it may be difficult to satisfy the requirement that $\tilde\delta+\delta=0$.  For spatially uniform fields, it may be possible to cancel a non-zero Zeeman light shift $\tilde\delta$ over the entire area of interaction by a suitable choice of the two-photon detuning $\delta$.  For a multimode interaction where the relative intensities of the laser fields is allowed to vary spatially, this is in general not experimentally feasible.  As a result, in the analysis that follows we will assume that $\delta$ is zero, and when coupling atoms between the states we will require the ratio of laser intensities to be such that $\tilde\delta$ is zero, i.e. 

\begin{equation}\label{eq:thetaconst2}
(\chi_1^a-\chi_2^{bc})I_-+(\chi_1^{bc}-\chi_2^d)I_+=0
\end{equation}

Solving \eqref{eq:thetaconst1} and \eqref{eq:thetaconst2} for $I_+$ and $I_-$ yields

\begin{equation}\label{eq:I+theta}
I_+^{(\theta)}=\frac{\sqrt{\chi_1^a-\chi_2^{bc} }}{\eta_{12}^{bc}\sqrt{\chi_2^d-\chi_1^{bc} }}\frac{\theta}{\tau}
\end{equation}

\begin{equation}\label{eq:I-theta}
I_-^{(\theta)}=\frac{\sqrt{\chi_2^d-\chi_1^{bc} }}{\eta_{12}^{bc}\sqrt{\chi_1^a-\chi_2^{bc} }}\frac{\theta}{\tau}
\end{equation}

\begin{figure}[ht]
	\centering
		\includegraphics[scale=1.1]{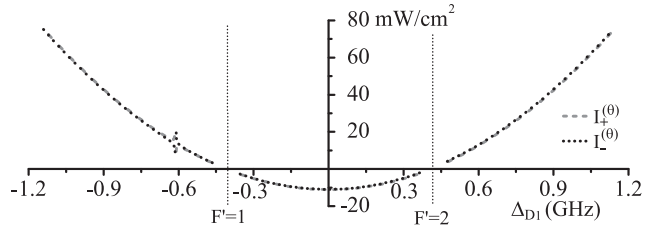}
	\caption{Detuning dependence of the intensities required to cause a 180$^\circ$ $\theta$ rotation of the  pseudospin vector describing the Raman-coupled $\left|F=1,m_F=-1\right\rangle$ and $\left|F=1,m_F=1\right\rangle$ states, subject to the condition that $\tilde\delta=0$.  Negative intensities correspond to a sign change in the direction of rotation of $\theta$.  Pulse duration $\tau=20$ $\mu$s, magnetic bias field B = 1.33 Gauss}
	\label{fig:F1theta}
\end{figure}

The relations defined in Eqns. \eqref{eq:I+theta} and \eqref{eq:I-theta} are plotted in Fig.~\ref{fig:F1theta} as a function of detuning, for typical experimental values of pulse duration and magnetic bias field.  It should be noted that for detunings between the excited states, the sense of rotation of the pseudospin vector reverses, which is indicated in the plot by negative intensities.  The small feature near -620 MHz occurs due to the fact that the Zeeman light shifts for the $\sigma^+$ and $\sigma^-$ fields vanish there at detunings which are not quite identical for B $\neq$ 0, making it impossible to satisfy the requirement that $\tilde\delta=0$ in the neighborhood of those points.

From Fig.~\ref{fig:F1theta}, it is apparent that using nearly equal intensities in the $\sigma^+$ and $\sigma^-$ fields satisfies Eq. \eqref{eq:thetaconst2} over essentially the entire range of detunings.  This is a consequence of the high degree of symmetry in the dipole matrix elements and state linkages.  We will show later that in systems with less symmetry, the solutions will not generally be so simple. 

Having determined how to cause a $\theta$ rotation of the pseudospin vector, we now consider the problem of controlling $\phi$, or in other words, the relative phase of the two states.  Applying only the $\sigma^+$ or the $\sigma^-$ laser fields causes no change in the relative population of the states, but will cause a change in their relative phases if the Zeeman light shift is nonzero.  As discussed above, this effect can be understood in the pseudospin representation as being due to interaction with a fictitious magnetic field along the polar axis which causes the pseudospin vector to precess azimuthally at a rate proportional to the magnitude of the Zeeman light shift.  The intensity required to cause a given rotation in $\phi$ by applying the $\sigma^+$ or $\sigma^-$ polarized beams individually for a pulse of duration $\tau$ can be written as:

\begin{equation}\label{eq:Iphi+}
I_+^{(\phi)}=\frac{\phi/\tau }{\chi_2^d-\chi_1^{bc} }
\end{equation}

\begin{equation}\label{eq:Iphi-}
I_-^{(\phi)}=\frac{\phi/\tau }{\chi_1^a-\chi_2^{bc} }
\end{equation}

\begin{figure}[ht]
	\centering
		\includegraphics[scale=1]{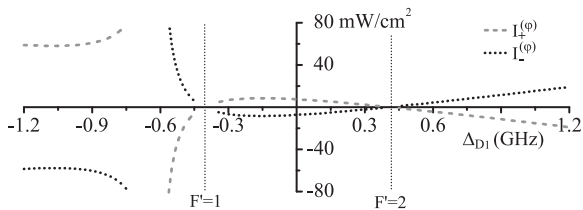}
	\caption{Detuning dependence of the intensities required to cause a 180$^\circ$ $\phi$ rotation of the  pseudospin vector describing the Raman-coupled $\left|F=1,m_F=-1\right\rangle$ and $\left|F=1,m_F=1\right\rangle$ states.  Negative intensities correspond to a sign change in the direction of rotation of $\phi$. $\tau$ = 20$\mu$s, B = 1.33 Gauss}
	\label{fig:F1phi}
\end{figure}

As with the expressions for rotation in $\theta$, the sign of the rotation is determined by the sign of the light shift coefficients, which can be either positive or negative.  For the $F=1$ Hamiltonian under consideration in this Section, the light shifts due to the $\sigma^+$ and $\sigma^-$ laser fields are opposite in sign and of nearly equal magnitude (see Fig.~\ref{fig:F1phi}).  As noted above, the Zeeman light shift from both laser fields vanishes near -620 MHz, which makes it impossible to cause any $\phi$ rotation with either laser field near that detuning. This is reflected in the plot by the obvious presence of the large singularity.  It should be noted that only the Zeeman light shift vanishes at this detuning, the scalar light shift is still nonzero.

The expressions (\ref{eq:I+theta}-\ref{eq:Iphi-}) plotted in Fig.~\ref{fig:F1theta} and \ref{fig:F1phi} give a clear picture of the experimental parameter space in which we must work to control the system with the Raman laser pulses.  There is a wide range of valid detunings over which we have a protocol for determining what optical pulses will cause a desired change of the state vector.  We will show how we have been able to apply this information experimentally to control the spinor wavefunction of an $F=1$ BEC in the experimental results presented in section \ref{sec:results}.

\subsection{ Application to the $^{87}$Rb $F=2$ Manifold}\label{ssec:F=2}

\begin{figure}[ht]
	\centering
		\includegraphics[scale=1]{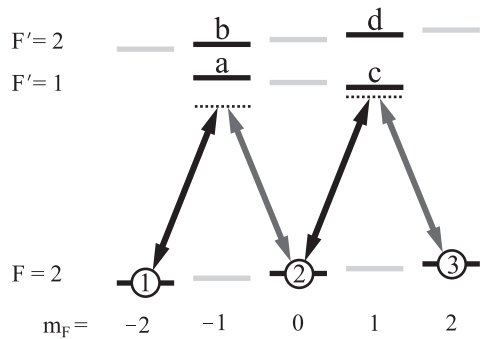}
	\caption{State linkage diagram for $(\sigma^+,\sigma^-)$ Raman coupling of the $F=2$ ground state manifold of $^{87}$Rb via the D$_1$ transitions. }
	\label{fig:F2levels}
\end{figure}

Having demonstrated the basic derivation of a protocol for the $F=1$ manifold, we turn to a more complicated case, which is to develop a useful model of Raman interactions within the $F=2$ ground state Zeeman manifold.  Several differences are immediately apparent, one being that there are two subgroups of the manifold which can be coupled by a $\sigma^+$, $\sigma^-$ beam pair: the states with $m_F=-2,0,2$ and those with $m_F=-1,1$.  Because the $m_F=2$ state is the only one in which it is possible to produce a BEC in a magnetic trap, we will limit our attention to the $m_F=-2,0,2$ subset, shown in Fig.~\ref{fig:F2levels}.  Constructing a Hamiltonian for these states in the interaction picture, making the rotating-wave approximation, and adiabatically eliminating the excited states as explained above results in the following effective 3-level Hamiltonian.

\begin{equation*}\label{eq:AEHF=2b}
\hbar
\left[ \begin{array}{ccc} 
\delta_{12}+\chi_1^{ab} I_{+}
&  
\eta_{12}^{ab}\sqrt{I_-I_+}e^{-i\xi_{\pm}}  
&
0                       
\\
\\
\eta_{12}^{ab}\sqrt{I_-I_+} e^{i\xi_{\pm}}
& 
\chi_2^{ab} I_{-} + \chi_2^{cd} I_{+}
&
\eta_{23}^{cd}\sqrt{I_-I_+} e^{-i\xi_{\pm}}
\\
\\
0
&
\eta_{23}^{cd}\sqrt{I_-I_+} e^{i\xi_{\pm}}
&
\delta_{23}+\chi_3^{cd} I_{-}
\end{array}\right]
\end{equation*}

As in the previous example of the $F=1$ manifold, here we see light shift terms ($\chi_i^{\alpha\beta}$) in the diagonal elements and two-photon coupling terms ($\eta_{ij}^{\alpha\beta}$) in the off-diagonal elements.  The definition of these parameters $\eta$ and $\chi$ can be inferred from the sub- and superscripts by noting the state labels indicated in Fig.~\ref{fig:F2levels}, and following the pattern of the definitions shown for the $F=1$ manifold (Eqns. \ref{eq:F1chis}, \ref{eq:F1eta}).  Several differences in form are immediately apparent, such as the separately defined two-photon detunings for the 1$\leftrightarrow$2 and 2$\leftrightarrow$3 transitions.

\begin{align*}
\delta_{12}&=(E_1-E_2)/\hbar+(\omega_+-\omega_-)
    \\
\delta_{23}&=(E_2-E_3)/\hbar+(\omega_+-\omega_-)
\end{align*}

For $ B\approx0 $, the energies of the Zeeman sublevels within a manifold are degenerate, and in this limit the interaction terms on both sides of the``M'' are equal, i.e. $\delta_{12}=\delta_{13}$, and $\eta_{12}^{ab}=\eta_{23}^{cd}$.  Although it is possible to adopt a pseudospin-1 representation of the system in this limit by expressing the Hamiltonian in terms of a suitable basis of $3\times3$ matrices, we will consider here a different approach.  In the presence of a magnetic field, the nonlinear part of the Zeeman energy shifts breaks the degeneracy of the two-photon transitions, i.e. $\delta_{12}\neq\delta_{23}$.  For low to moderate fields ($<$100 G), this relative shift of the two transitions frequencies increases quadratically with the magnetic field, by 575 Hz/G$^2$.  For sufficiently large field and sufficiently small optical pulse bandwidth, the 1$\leftrightarrow$2 and 2$\leftrightarrow$3 transitions can be treated as separately addressable pseudospin-1/2 subsystems with Hamiltonians similar to Eq. \eqref{eq:F1spinham}, but with the interaction parameters for the 1$\leftrightarrow$2 transition given by

\begin{equation}\label{eq:adO12}
\begin{split}
\tilde\alpha_{12}&=I_+(\chi_1^{ab}+\chi_2^{cd}) +I_-\chi_2^{ab}\\
\tilde\delta_{12}&=I_+(\chi_1^{ab}-\chi_2^{cd}) -I_-\chi_2^{ab}\\
\tilde\Omega_{12}&=2\eta_{12}^{ab}\sqrt{I_+I_-}
\end{split}
\end{equation}

\noindent and the parameters for the $2\leftrightarrow3$ transition given by
\begin{equation}\label{eq:adO23}
\begin{split}
\tilde\alpha_{23}&=I_-(\chi_2^{ab}+\chi_3^{cd})+I_+\chi_2^{cd}\\       
\tilde\delta_{23}&=I_-(\chi_2^{ab}-\chi_3^{cd})+I_+\chi_2^{cd}\\
\tilde\Omega_{23}&=2\,\eta_{23}^{cd}\sqrt{I_+I_-} 
\end{split}
\end{equation}

\noindent As before, $\tilde\alpha$ is a scalar light shift, $\tilde\delta$ is a Zeeman light shift, and $\tilde\Omega$ is the effective two-photon Rabi frequency for the given transition. It is important to note that these pseudospin-1/2 subsystems of the $F=2$ manifold are inherently quite asymmetric, which has a significant effect on the overall nature of the response of the system to the applied laser fields. 

\begin{figure}[ht]
	\centering
		\includegraphics[scale=1.1]{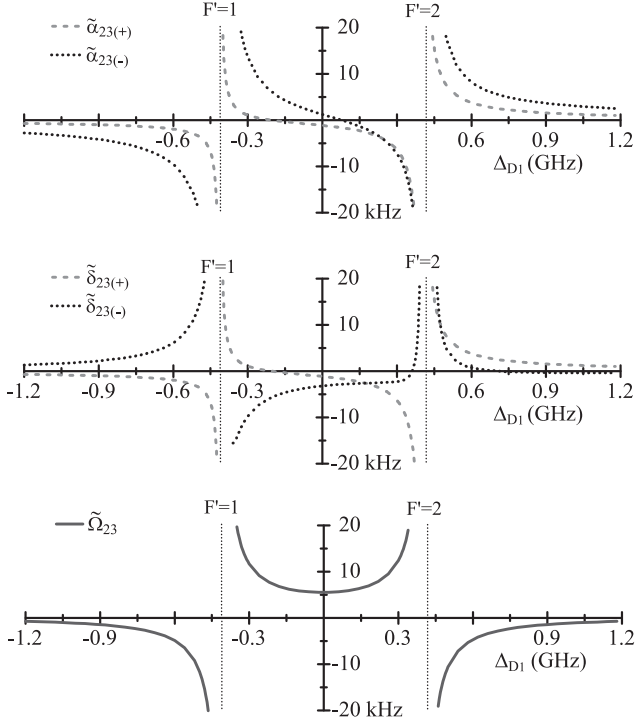}
	\caption{Detuning dependence of the scalar light shift $\tilde\alpha_{23}$, Zeeman light shift $\tilde\delta_{23}$, and two-photon Rabi frequency $\tilde\Omega_{23}$ associated with the 2$\leftrightarrow$3 subspace of the $F=2$ manifold.  The scalar and Zeeman light shifts are expressed in terms of their components which depend on the intensities of the $\sigma^+$ and $\sigma^-$ polarized laser fields. ($I_+,I_-=1$ mW/cm$^2$, B = 17 Gauss)}
	\label{fig:alphadeltaOmega23}
\end{figure}

The dependence of the parameters for the 2$\leftrightarrow$3 subspace (Eq.~\ref{eq:adO23}) on single-photon detuning is shown in Fig.~\ref{fig:alphadeltaOmega23}.  The  magnetic bias field assumed for these plots is relatively large, which results in a split in the degeneracy of the two-photon transitions significantly greater than our typical pulse bandwidth.  Although the variation of $\tilde\alpha$ and $\tilde\Omega$ with detuning are generally unremarkable in comparison with the plots for the $F=1$ system, the Zeeman light shifts $\tilde\delta_+$ and $\tilde\delta_-$  are clearly not equal and opposite in magnitude, as they were for the $F=1$ system.  This has important ramifications for our attempts to produce a protocol for controlling this system.  

\begin{figure}[ht]
	\centering
		\includegraphics[scale=1.1]{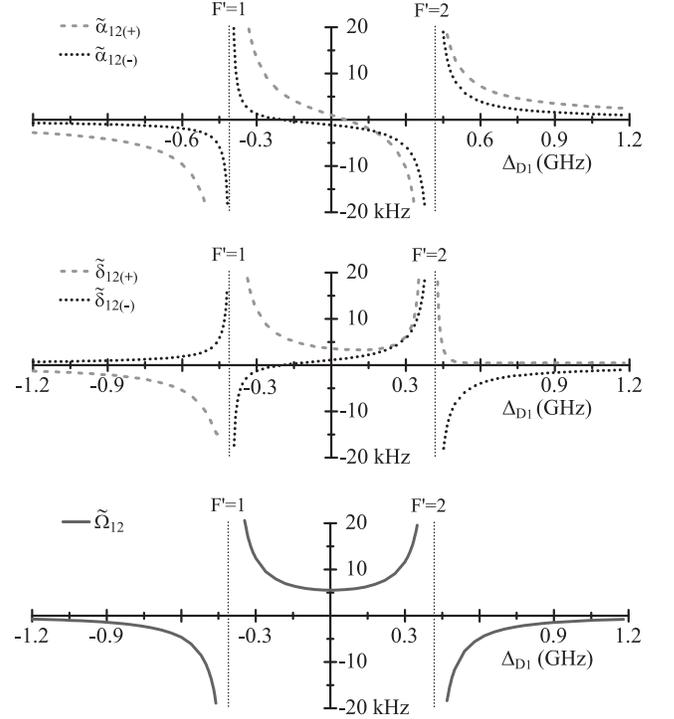}
	\caption{Detuning dependence of the scalar light shift $\tilde\alpha_{12}$, Zeeman light shift $\tilde\delta_{12}$, and two-photon Rabi frequency $\tilde\Omega_{12}$ associated with the 1$\leftrightarrow$2 subspace of the $F=2$ manifold.  The scalar and Zeeman light shifts are expressed in terms of their components which depend on the intensities of the $\sigma^+$ and $\sigma^-$ polarized laser fields. ($I_+,I_-=1$ mW/cm$^2$, B = 17 Gauss)}
	\label{fig:alphadeltaOmega12}
\end{figure}

At zero magnetic bias field it is important to note that the plots of parameters \eqref{eq:adO12} and \eqref{eq:adO23}, for the 1$\leftrightarrow$2 and 2$\leftrightarrow$3 subspaces, respectively, are essentially identical if the identity of the $\sigma^+$ and $\sigma^-$ fields is interchanged.  For magnetic fields of more than a few Gauss, the asymmetry of the (non-fictitious) Zeeman shift causes significant differences between the two subspaces as the field is increased, particularly for detunings above resonance with the $F'=2$ manifold. This can be observed in the difference between Figs. \ref{fig:alphadeltaOmega23} and \ref{fig:alphadeltaOmega12}.  The light shifts are much less dependent on magnetic field and in several ways much more well-behaved when the beams are detuned below $F'=1$, which makes it a somewhat more attractive region to work in experimentally.

The expressions \eqref{eq:adO12} and \eqref{eq:adO23} allow us to similarly transform these subspaces into a pseudospin representation and think in terms of fictitious fields to cause the system to evolve as desired.  Repeating the approach used to generate a protocol for the $F=1$ manifold produces a similar set of relations indicating how to cause a given $\theta$ or $\phi$ rotation of the pseudospin vector within the subspace.  For the 2$\leftrightarrow$3 transition these are:

\begin{equation*}
I_+^{(\theta)}=
\frac{ \sqrt{\chi_2^{ab}-\chi_3^{cd}} }{\eta_{23}^{cd}\sqrt{\chi_2^{cd} } }
\frac{\theta}{\tau}; \ \ \ 
I_-^{(\theta)}=
\frac{ \sqrt{\chi_2^{cd}} }{\eta_{23}^{cd}\sqrt{\chi_2^{ab}-\chi_3^{cd} } }
\frac{\theta}{\tau}
\end{equation*}

\begin{equation*}
I_+^{(\phi)}=\frac{\phi/\tau }{\chi_2^{cd}}; \ \ \ \ 
I_-^{(\phi)}=\frac{\phi/\tau }{\chi_2^{ab}-\chi_3^{cd} }
\end{equation*}

\begin{figure}[ht]
	\centering
		\includegraphics[scale=1.0]{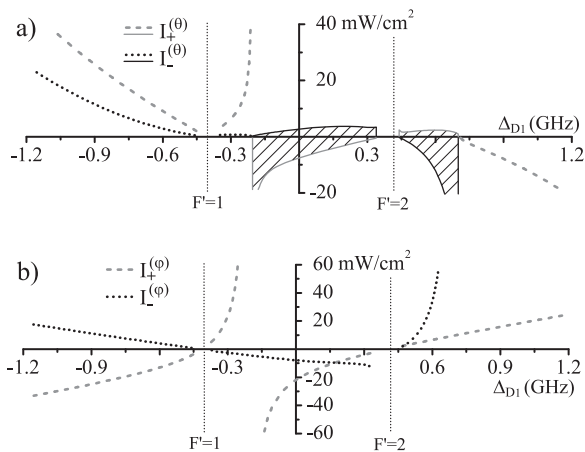}
	\caption{Plot of the intensities required to cause a 180$^\circ$ $\theta$ rotation (a) or $\phi$ rotation (b) of the pseudospin vector describing the Raman-coupled $\left|F=2,m_F=0\right\rangle$ and $\left|F=2,m_F=2\right\rangle$ states, subject to the condition that $\tilde\delta_{23}=0$.  Negative intensities correspond to a sign change in the direction of rotation of $\theta$. Cross-hatching indicates imaginary-valued, nonphysical solutions. Pulse duration $\tau=20$ $\mu$s, magnetic bias field B = 17 Gauss}
	\label{fig:F2thetaphi23}
\end{figure}

The expressions governing the evolution of the 2$\leftrightarrow$3 subspace in $\theta$ are plotted as a function of detuning in Fig.~\ref{fig:F2thetaphi23}(a).  The most noticeable feature of this plot is the large region for which the solution is not real-valued, and hence unphysical.  Over that range of detunings $\tilde\delta$ cannot be made zero because the sign of the Zeeman light shift for the $\sigma^+$ and $\sigma^-$ laser fields is the same, and they cannot be made to cancel each other out. It should be noted that coherent transfer is still possible in this range of detunings, but only if the real two-photon detuning is adjusted such that it compensates for the Zeeman light shift. 

Another noticeable feature of Fig.~\ref{fig:F2thetaphi23}(a) is the large variation in the ratio of intensities required to ensure that $\tilde\delta=0$. This can be understood as a consequence of the large asymmetry in the dipole matrix elements for the transitions involved, and the fact that varying the detuning changes the strength of coupling to the different excited state manifolds. 

Certainly for the $F=2$ manifold, choosing an appropriate laser detuning is an important factor in being able to exert control over the state of the BEC.  We have in general had the most reliable experimental results with an $F=2$ BEC when operating at a detuning several hundred MHz below resonance with the $F'=1$ manifold. Alternate choices of detuning are possible, but generally more challenging to implement experimentally.  Some examples of our approach to control of the BEC will be shown in the next section.

\section{ Experimental Results }\label{sec:results}

We have performed a number of experimental tests of the protocols derived in the preceding sections.  The experimental procedure common to all of the experiments is as follows. 

A BEC of $\approx10^5$ atoms of $^{87}$Rb is prepared in a magnetic trap, spin-polarized in either the  $\left|F=1,m_F=-1\right\rangle$ or the $\left|F=2,m_F=2\right\rangle$ state.   This BEC is then released from the magnetic trap and allowed to expand for 9 ms to a diameter of $\approx70$ $\mu$m, in the presence of a weak magnetic bias field. The average density of the BEC is then $\approx5\times10^{11}$atoms/cm$^3$, and the mean field energy shift is on the order of a few tens of Hertz, which is experimentally negligible.  Once this target is prepared, the Raman beams are pulsed on during an interaction time lasting 5-30 $\mu$s, depending on the experiment.  The diameter of the Raman beams is large enough (1.7 mm) to ensure that the beam intensity is essentially uniform over the area of the BEC, with $\Delta I/I_{avg}<0.01$. The magnetic bias field over this region is uniform to better than 1 mG at the maximum applied field strength of 17 G. 

 After the laser interaction, we determine the final distribution of population among the Zeeman sublevels by applying a strong transverse magnetic field gradient of $\approx\,$300 G/cm for 1 ms, which causes a Stern-Gerlach separation of the different spin components during a subsequent 20 ms time-of-flight \cite{StengerSpinN98}.  The field gradient is uniform to within $\pm$10\% over the condensate, which is sufficient to avoid significant distortion due to magnetic lensing during the separation process. Once separated, the clouds are imaged using typical absorption imaging techniques. The population in each Zeeman sublevel is quantified by integrating the measured density over the appropriate regions of the image, correcting for variations in background level.  The ratio of population in the target state to the total population can then be calculated and compared to theoretical predictions.  The details of the individual experimental tests we conducted are explained below.

\subsection{Rabi Oscillations}

The primary test of the model and the protocols we have derived is to see whether complete population oscillation occurs at the frequency predicted for the specified intensities of the $\sigma^+$ and $\sigma^-$ fields. Although the typical method of observing Rabi oscillations is to apply a field and vary the duration of interactions, this approach is not ideal for testing of our model of the system.

 The approach we chose to use to observe Rabi oscillations in our system was to hold the pulse duration constant and increase the intensities in the ratio indicated by the equations, observing the resulting change in the target state population.  Any mismatch between theory and experiment would result in the system being driven increasingly out of resonance at higher pulse intensities by the increasing light shifts.  Another motivation to perform such a test is that the model was developed with the intention to apply it to laser modes with spatially varying intensity.  The performance of the model in this respect is therefore an important metric of its utility.  

\begin{figure}[ht]
	\centering
		\includegraphics[scale=1.1]{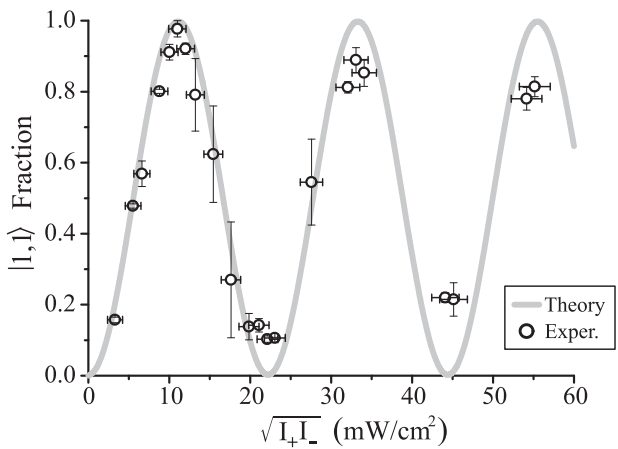}
	\caption{Population in the $\left|1,1\right\rangle$ state after a 20 $\mu$s Raman laser pulse as a function of square root of the product of the beam intensities. The vertical error bars indicate the scatter range in the measurements, and the horizontal error bars indicate the uncertainty in the beam intensity due to laser power fluctuations.  The grey line is the theoretically predicted intensity dependence.}
	\label{fig:F1ampexp}
\end{figure}

The results of an experiment demonstrating Rabi oscillations between the $\left|1,-1\right\rangle$ and $\left|1,1\right\rangle$ states are shown in Fig.~\ref{fig:F1ampexp}.  In this experiment, we used optical pulses of 20 $\mu$s duration and intensities of up to 55 mW/cm$^2$. The beams were detuned midway between the $F'=1$ and $F'=2$ transitions, i.e. at $\Delta_{D_1}=0$ on the plots of section \ref{sec:modeling}.  At this detuning, the ratio of I$_+$/I$_-$ which makes $\tilde\delta=0$ is 1.01.  The magnetic bias field applied to the system was 1.33 Gauss, resulting in a ground-state Zeeman splitting of 0.93 MHz.  
  
  The plot shows the fraction of the population transferred to the $\left|1,1\right\rangle$ state as a function of the square root of the product of the beam intensities, which is proportional to $\tilde\Omega$.  The light grey line is the theoretical prediction from the model.  Each data point is the average of several runs of the experiment, with the error bars indicating the full range of scatter of the individual runs. Due to beam intensity variations, the vertical scatter is more pronounced in the regions where the function is steeply varying, however the model and the data are in excellent agreement. 

\begin{figure}[ht]
	\centering
		\includegraphics[scale=1.0]{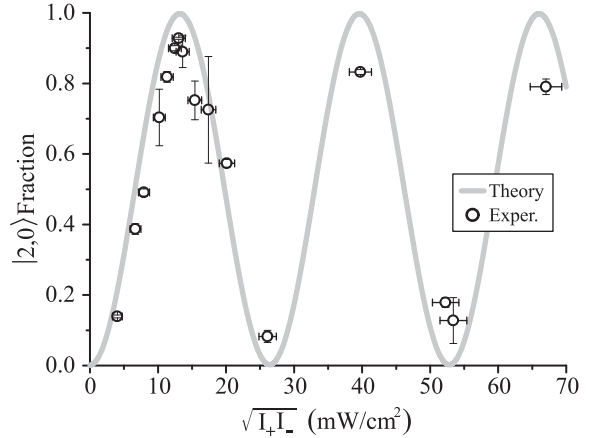}
	\caption{ Population in the $\left|2,0\right\rangle$ state after a 20 $\mu$s Raman laser pulse as a function of square root of the product of the beam intensities. The vertical error bars indicate the scatter range in the measurements, and the horizontal error bars indicate the uncertainty in the beam intensity due to laser power fluctuations}
	\label{fig:F2ampexp}
\end{figure}

The results from a similar experiment demonstrating coherent oscillations between the $\left|2,2\right\rangle$ state and the $\left|2,0\right\rangle$ state are shown in Fig.~\ref{fig:F2ampexp}.  The magnetic bias field applied in this experiment was 17 Gauss, resulting in a ground-state Zeeman splitting of 12 MHz, and a difference between the two-photon resonances of $\delta_{12}-\delta_{23}$ = 166 kHz. This splitting was large enough that with laser pulses of 20 $\mu$s duration (1/$\tau$=50 kHz), we were able to address the $\left|2,2\right\rangle\leftrightarrow\left|2,0\right\rangle$ transition separately from the $\left|2,-2\right\rangle\leftrightarrow\left|2,0\right\rangle$ transition.

For this experiment, $\Delta_{D_1}$ was set to -0.8 GHz, below resonance with the $F'=1$ excited state manifold. We maintained the ratio of laser intensities at $I_+/I_-=2.40$, which is the value that the model predicts will result in $\tilde\delta=0$. It is worthwhile to note that there is an upper limit to the intensities which can be used in these experiments, because of the necessity of avoiding incoherent exitation of the atoms.  For the detunings and pulse durations used in these experiments, we typically only begin to see significant signs of incoherent excitation at intensities well above 200 mW/cm$^2$.

In both Fig.~\ref{fig:F1ampexp} and  \ref{fig:F2ampexp}, the experimental results are in excellent agreement with the theoretical predictions.   We emphasize that the same calibration values for the absolute intensity have been used for all the data sets presented in this Article, and there are no free parameters in any of the theoretical curves.  We do note that there is some decrease in the amplitude of the oscillation at higher intensities.  This could be due either to technical factors such as fluctuations in the beam intensities and drift in the magnetic bias field, or possibly small physical influences neglected in our current model, such as beam propagation effects.  If the cause is one of the latter, it would prove to be an interesting  subject for future study. 
 
\subsection{Phase Control in the $F=1$ Manifold}

After having established that we can cause several complete Rabi oscillations between two Zeeman sublevels by applying an appropriately configured Raman pulse pair, we can approach more sophisticated problems in system control.  A more demanding test of the protocols derived in Section \ref{sec:modeling} is to attempt to demonstrate control of the relative phase of the states being coupled.  

\begin{figure}[ht]
	\centering
		\includegraphics[scale=1.1]{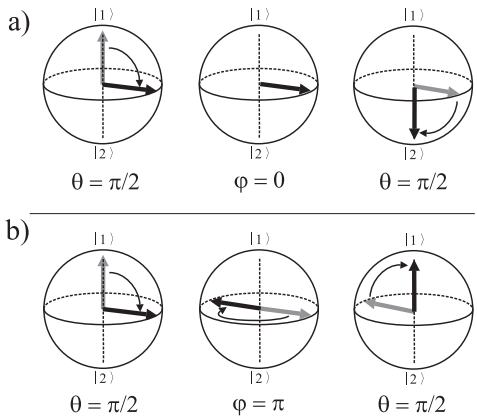}
	\caption{ Applying two successive $\theta=\pi/2$ pulses with no phase shift results in complete transfer to the final state (a). Turning on only one of the Raman lasers causes a phase shift, rotating the pseudospin vector in $\phi$.  For a $\phi=\pi$ rotation (b), the population returns to the initial state.}
	\label{fig:Bloch}
\end{figure}

One simple way of showing phase control in these pseudospin-1/2 systems is to perform a Ramsey-fringe type experiment, using a three step pulse sequence as depicted in Fig.~\ref{fig:Bloch}.  The first step in this sequence is to apply a $\sigma^+$-$\sigma^-$ pulse pair which causes the pseudospin vector to precess 90$^\circ$  in the $\theta$ direction ($\pi/2$ effective pulse area).  The second step is to apply only one or the other of the laser fields, which causes the pseudospin to precess in $\phi$ by an amount which depends on the magnitude of the Zeeman light shift and the duration of the pulse.  The last step is to again apply a $\theta=\pi/2$ pulse.   If neither of the fields is turned on in between the two $\theta=\pi/2$ pulses, after the final step all the population should end up in the target state. (Fig.~\ref{fig:Bloch}a)  If a small phase shift is applied during the second step, the last operation will not move all of the population into the final state.  If the $\phi$ phase shift is 180$^\circ$ ($\pi$), the action of the final pulse will be to actually return all of the population to the initial state. (Fig.~\ref{fig:Bloch}b)

\begin{figure}[ht]
	\centering
		\includegraphics[scale=1.1]{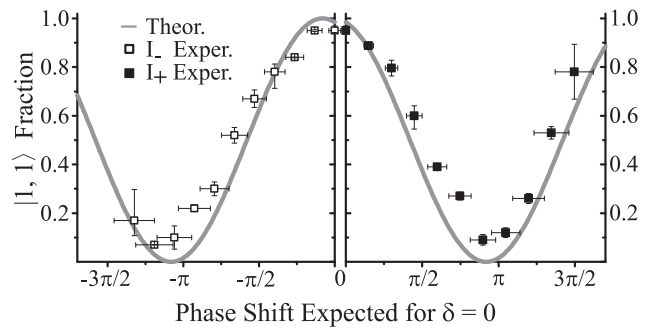}
	\caption{ Experimental results of applying the $\theta=\pi/2$, $\phi$,  $\theta=\pi/2$ pulse sequence described above to the $\left|1,-1\right\rangle\:\leftrightarrow\:\left|1,1\right\rangle$ pseudospin system.  Application of a $\phi$ changing pulse is shown to cause the expected oscillation in the final state population.    The grey line is a theoretical prediction taking into account a nonzero $\tilde\delta$ caused by lowering the $\sigma_-$ beam intensity by 5\%.}	\label{fig:F1phaseexp}
\end{figure}

  The results of an experiment implementing the pulse sequence of Fig.~\ref{fig:Bloch} are shown in Fig.~\ref{fig:F1phaseexp}.  For this experiment, we used the $\left|1,-1\right\rangle\:\leftrightarrow\:\left|1,1\right\rangle$ system, with the laser detunings set to $\Delta_{D_1}=0$ as in the coherent oscillation experiment of Fig.~\ref{fig:F1ampexp} above. The $\theta=\pi/2$ pulses were 5 $\mu$s in duration, separated by 10 $\mu$s.  After the first $\theta=\pi/2$ pulse, we left on either the $\sigma^+$ or the $\sigma^-$ laser field for a duration of 0-10 $\mu$s in order to cause the desired phase shift. For the data shown, the ratio of $I_+/I_-$ was offset by 5\% from the values which would make $\tilde\delta=0$, in order to make the sign difference in the rotation caused by the $\sigma^+$ and $\sigma^-$ beam in this configuration explicitly apparent.  The theoretical curve in the figure has been shifted accordingly, and is in reasonable agreement with the experimental data.  This confirms that phase control of the system is indeed possible, and occurs as predicted by the model. 

\subsection{Two-photon Lineshape in the $F=2$ Manifold}\label{ssec:2pdet}

In the previous Sections we have focused on Raman-coupling of two selected states out of a Zeeman manifold.  The distinction made in Section \ref{ssec:F=2} between treating the $\left|2,2\right\rangle$, $\left|2,0\right\rangle$, and $\left|2,-2\right\rangle$ states as either a pseudospin-1 system at zero magnetic field, or a pair of pseudospin-1/2 systems in the presence of a large magnetic field is, of course, an oversimplification.  At moderate magnetic fields, using pulses of sufficient bandwidth, this trio of states in the $F=2$ manifold can be coupled simultaneously even though the two-photon transitions are not quite degenerate.  This allows for the intriguing possibility of using a single laser pulse pair to distribute population into all three states, where the fraction of the population in each state and their relative phase can be controlled by varying the splitting between $\delta_{12}$ and $\delta_{23}$ with the magnetic field, and adjusting the two-photon detuning with respect to these resonances in a manner determined by numerical modeling of the system.

\begin{figure}[ht]
	\centering
		\includegraphics[scale=1.2]{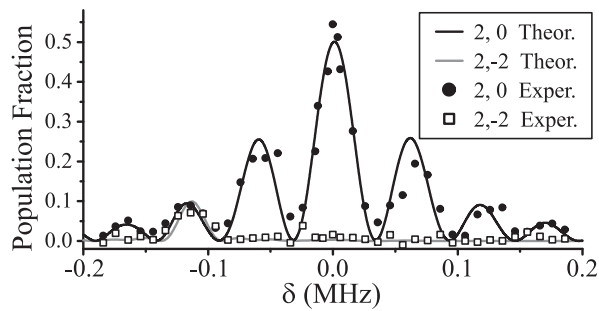}
	\caption{ Fractional population transferred to $\left|2,0\right\rangle$ and $\left|2,-2\right\rangle$ from a BEC initially in the $\left|2,2\right\rangle$ state by a 20 $\mu$s Raman laser pulse for different choices of the two-photon detuning $\delta$.  The magnitude and ratio of the intensities were set for an effective $\theta=1.5\pi$ pulse area on two-photon resonance with the $\left|2,2\right\rangle\:\leftrightarrow\:\left|2,0\right\rangle$ transition, with the detuning of the beams set at $\Delta_{D_1}$ = -800 MHz.  The applied magnetic bias field was 17 Gauss, for which $\delta_{12}-\delta_{23}$=166 kHz.}
	\label{fig:2PDplot}
\end{figure}

The results of an experiment demonstrating the two-photon detuning dependence of the fractional transfer to the $\left|2,0\right\rangle$ and $\left|2,-2\right\rangle$ states are shown in Fig.~\ref{fig:2PDplot}.  The applied magnetic field used for that data set is still relatively large, such that the energy separation of the two photon resonances is several times the pulse bandwidth.  The intensities used were I$_+$= 30.6 mW/cm$^2$ and I$_-$= 12.8 mW/cm$^2$, which are appropriate to make $\tilde\delta_{23}=0$, with an effective two-photon pulse area of 1.5 $\pi$ for resonance with that transition. It should be noted that in this plot the maximum vertical scale for the transfer fraction is consequently 0.5, not 1. With the system in this configuration, two-photon detuning was varied about resonance with the $\left|2,2\right\rangle$ to $\left|2,0\right\rangle$ transition by $\pm$200 MHz, and the fractional transfer of population to the $\left|2,0\right\rangle$ and $\left|2,-2\right\rangle$ states was recorded, taking the average of several experimental runs.  The oscillatory behavior of the transfer to $\left|2,0\right\rangle$ appears with a periodicity close to that expected from the pulse bandwidth.  As the two photon detuning is moved below -100 MHz, however the lasers begin to come into resonance with the transition to the $\left|2,-2\right\rangle$ state, and population begins to appear there as well.  The solid curves appearing behind the data points are from a numerical model of the three-state dynamics in this configuration, and can be seen to be in excellent agreement with the experimental results.
  
\begin{figure}[ht]
	\centering
		\includegraphics[scale=1.0]{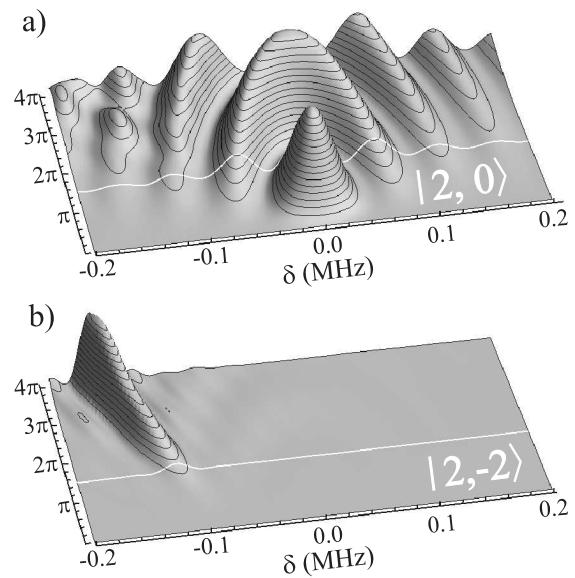}
	\caption{ Numerical prediction of the population transferred to $\left|2,0\right\rangle$ (a) and $\left|2,-2\right\rangle$ (b) from a BEC initially in the $\left|2,2\right\rangle$ state, as a function of two-photon detuning and effective pulse area. The ratio of the intensities was set to eliminate $\tilde\delta$ for the $\left|2,2\right\rangle\:\leftrightarrow\:\left|2,0\right\rangle$ transition.  $\Delta_{D_1}$ = -800 MHz, B = 17 Gauss, for which $\delta_{12}-\delta_{23}$=166 kHz. The theoretical curves shown in Fig.~\ref{fig:2PDplot} are lineouts from these plots at the indicated location of $\theta=1.5\pi$ The highest peaks corresponds to complete population transfer.}
	\label{fig:tpd3D}
\end{figure}

The general dependence of this three-state coupling on beam intensity adds another degree of complexity to the picture, as can be seen in the numerical predictions plotted in Fig.~\ref{fig:tpd3D}.  The upper plot (a) indicates the fractional transfer to the $\left|2,0\right\rangle$ state as a function of two-photon detuning in the broad axis, and as a function of the effective pulse area when resonant with the $\left|2,2\right\rangle\:\leftrightarrow\:\left|2,0\right\rangle$ transition along the other axis.  The intensity ratio and magnetic field are the same as those used in the results of Fig.~\ref{fig:2PDplot}.  The lower plot (b) shows the transfer fraction to the $\left|2,-2\right\rangle$ state for the same experimental configuration. The theoretical curves shown in Fig.~\ref{fig:2PDplot} are a cross section taken from the data of Fig.~\ref{fig:tpd3D}, as indicated by the white lines on the surfaces. 

Several interesting features are apparent in Fig.~\ref{fig:tpd3D}, the most striking of which is that for high enough intensity and a two photon detuning of about $\delta_{12}$=-160kHz, all the population bypasses the $\left|2,0\right\rangle$ state and ends up in $\left|2,-2\right\rangle$.  This resonance ridge is noticeably narrower than the peaks seen in  Fig.~\ref{fig:tpd3D} a, and clearly moves off toward lower two-photon detuning as the beam intensities are increased.  These characteristics are a result of the increasing uncompensated Zeeman light shift of the $\left|2,0\right\rangle\leftrightarrow \left|2,-2\right\rangle$ subsystem.  The ability to place population in all three selected states in a tunable, deterministic manner is an extremely useful feature of the Raman-coupling system, allowing for the creation of more complex multi-component spin textures.
 
\section{Conclusion}
We have shown that the D$_1$ transitions in $^{87}$Rb can be used to coherently control the amplitude and phase of selected Zeeman sublevels in both the $F=1$ and $F=2$ ground state manifolds.  The model we have put forward accurately predicts the dynamics of the system and has allowed us to create simple protocols for producing a desired change in the atomic state. We have demonstrated successful implementation of these protocols both for single-pulse-pair operations, and for sequences of pulses. The results presented here are specifically applied to an untrapped, freely expanding BEC, however, this approach can be adapted to model Raman-coupling of a BEC in a trapping potential, especially if the pulse bandwidth is large compared to the frequency shifts due to the trapping potential and mean field.

Although we have derived and tested the model presented here in the plane-wave limit, it can be readily extended to describe interaction geometries employing laser fields with spatially-varying intensities and phases.  For example, it is possible to create vortices in the BEC order parameter by using optical vortex beams with an azimuthal phase winding, such as the Laguerre-Gaussian modes\cite{WrightOpticalPRA08}. This was the underlying motivation for this work, and we now have a clear physical picture of how to create complex, spatially-varying spin textures, such as two and three component coreless vortex states.  We anticipate that the principles and techniques presented here have the potential to greatly facilitate ongoing studies of topological states in quantum spin fluids.

This work was supported by NSF and ARO. LSL acknowledges support from the Laboratory for Laser Energetics.

\end{document}